\documentclass[10pt]{article}
\usepackage{graphicx}% Include figure files
\usepackage{dcolumn}% Align table columns on decimal point
\usepackage{bm}% bold math
\usepackage{color}% bold math
\usepackage{amssymb}
\usepackage{amsmath}
\usepackage{multicol,times,subfigure}
\usepackage{bm}
\usepackage{color}
\usepackage{url}
\usepackage{colortbl}
\usepackage{multirow}

\usepackage{enumitem}
\begin{document}

\title{\textbf{Identifying the Academic Rising Stars}}
\author{% \normalsize
Chuxu Zhang{\small$^{\mbox{1,2,3}}$},
 Chuang Liu{\small$^{\mbox{1}}$},
Lu Yu{\small$^{\mbox{5}}$},
Zi-Ke Zhang{\small$^{\mbox{1}}$}
 and Tao Zhou{\small$^{\mbox{3,4}}$}
}

\maketitle

\vspace{-5mm}

\noindent
$^1$Alibaba Research Center for Complexity Sciences, Hangzhou Normal University, Hangzhou, China.
$^2$Department of Computer Science, Rutgers University, NJ, USA.
$^3$Big Data Research Center, University of Electronic Science and Technology of China, Chengdu, China.
$^4$CompleX Lab, Web Sciences Center, University of Electronic Science and Technology of China, Chengdu, China.
$^5$Alibaba Group, Hangzhou, China.

\medskip
\noindent
{\small{Correspondence and requests for materials should be addressed to C.L. (liuchuang@hznu.edu.cn), Z.K.Z. (zhangzike@gmail.com) and T.Z. (zhutouster@gmail.com)}}

%\vspace{1mm}
%\begin{center}
%({\em{Received XXX; Accepted XXX; Published XXX}})
%\end{center}
%\vspace{1mm}

\begin{figure}[!t]%[htbp]
\begin{center}
\includegraphics[scale=0.65]{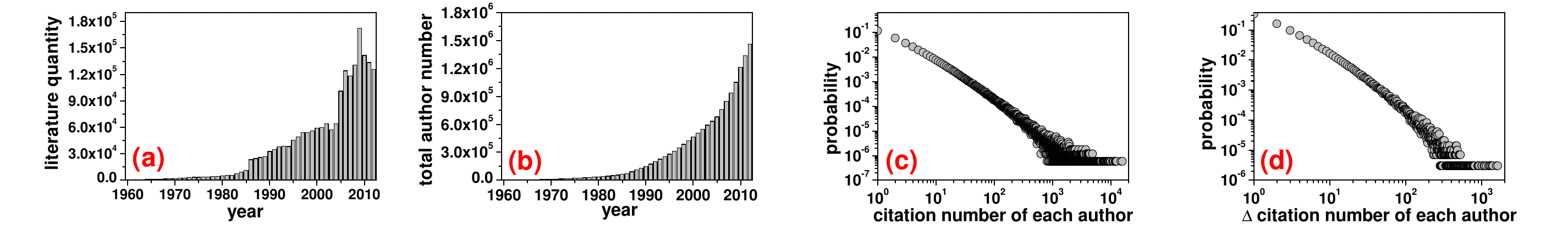}
\caption{(a) The volume of research literatures in each year. (b) The accumulative number of authors in each year from year 1960. (c) Distribution of all researchers' citation counts till year 2012. (d) Distribution of all researchers' citation increments from year 2008 to year 2012.}
\label{fig: statistic}
\end{center}
\end{figure}

%abstract
\begin{quote}
{\noindent \textbf{Predicting the fast-rising young researchers (Academic Rising Stars) in the future provides useful guidance to the research community, e.g., offering competitive candidates to university for young faculty hiring as they are expected to have success academic careers. In this work, given a set of young researchers who have published the first first-author paper recently, we solve the problem of how to effectively predict the  top $k\%$ researchers who achieve the highest citation increment in $\Delta t$ years. We explore a series of factors that can drive an author to be fast-rising and design a novel impact increment ranking learning (IIRL) algorithm that leverages those factors to predict the academic rising stars. Experimental results on the large ArnetMiner dataset with over 1.7 million authors demonstrate the effectiveness of IIRL. Specifically, it outperforms all given benchmark methods, with over 8\% average improvement. Further analysis demonstrates that the prediction models for different research topics follow the similar pattern. We also find that temporal features are the best indicators for rising stars prediction, while venue features are less relevant.}}
\end{quote}
%introduction
%\noindent{\textbf{Background.}} 
The growing scientific activities lead to the expanding body of literature, as well as the increasing academic population. Fig. \ref{fig: statistic}a and Fig. \ref{fig: statistic}b report the rapid increase of publication volume each year and the accumulative number of authors from 1960 in the ArnetMiner \footnote{\url{https://aminer.org/AMinerNetwork}} academic dataset of Computer Science \cite{tang2008arnetminer}. Despite the large number of researchers, their scientific impacts are heterogenous. Traditionally, the citation count is used as a measurement of  an author's scientific impact \cite{gehrke2003overview,castillo2007estimating,wang2013quantifying,shen2014modeling}. Fig. \ref{fig: statistic}c depicts the distribution of individual researcher' contribution in terms of the citation count, showing that less than 7\% of researchers have more than 20 citations. Meanwhile, the citation increasing trends of different authors are different. Fig. \ref{fig: statistic}d reports the citation increment distribution of all authors from year 2008 to 2012, showing that the distribution is power-law like and less than 10\% of authors have the fast-rising trend (each has the increment of citations being larger than 20) of scientific impact. An intuitive meaningful question arises that can we effectively predict the fast-rising ones (namely Academic Rising Stars - ARSes) among a set of young scholars who start academic research recently, as early identification of the ARSes may offer useful guidance to the research community like young faculty recruiting of university. 
%or assisting publication venue editor for contribution solicitation.
%\noindent{\textbf{Related Works.}} 

Generally, the question of ARSes prediction is closely related to scientific impact prediction, which has been extensively explored in recent years due to its importance in helping researchers to increase their reputations.
%Diverse factors attribute to the success of a researcher. For example, authors who conduct works with popular topic such as~\emph{Network Science} or publish papers in top venues such as \emph{Science/Nature}, generally attract much attention and may easily become influential.
Many previous works focused on predicting the future citation since the citation count prediction competition of 2003 KDD Cup \cite{gehrke2003overview}.
Castillo \emph{et al.} \cite{castillo2007estimating} estimated the paper's citation value
via using information about past articles written by the same author(s) of the paper. Yan \emph{et al.} \cite{yan2011citation,yan2012better} extracted a series of academic features (e.g., the author's citation number) and used several regression learning algorithms to predict the paper' citation number. Wang \emph{et al.} \cite{wang2013quantifying} and Shen \emph{et al.} \cite{shen2014modeling} revealed the underlying mechanism of scientific impact evolving, which can be used to predict citation value.
Besides citation number, the H-index \cite{hirsch2005index,hirsch2007does} proposed by Hirsch has been applied to measure both productivity and popularity of researchers. Dong \emph{et al.} \cite{dong2015collaboration} investigated the correlation between researchers' H-index values and their collaboration signatures. Dong \emph{et al.} \cite{dong2015will} extracted various academic features to classify the impactful papers that will contribute to improving the H-index. 
%Statistical regularities \cite{petersen2011statistical} of individual publication impact is proposed for better measurement of scientists' career progress. 
Despite the strong capability of previous works in predicting citation number, they are not effective for ARSes prediction. 
Because ARSes usually have limited accumulative citations while the fast citation rising helps them attract attentions from colleagues and have successful academic careers in the future. 

Unlike  predicting the exact citation value, in this work we focus on ranking the citation increments of different authors. 
Accordingly, we define the fast-rising researchers as the authors reach relatively large citation increments in a given time period. Our contributions are twofold: (i) we formalize the problem as a citation increment ranking task to identify the ARSes who rank in the front of others, (ii) we introduce a series of factors that are correlated with authors' future citation increments and design a novel ranking learning method to identify ARSes. The experiment results on a large academic dataset show the effectiveness of our proposed method. It outperforms all given baseline methods, with over 8\% average improvement. Besides citation count prediction, our work is related to 
%citation network analysis \cite{hunter2011dynamic} or
scientific network analysis \cite{newman2004coauthorship,tang2008arnetminer,newman2001scientific1,newman2001scientific2,newman2001structure,hunter2011dynamic}, 
credit allocation in academic collaboration \cite{shen2014collective,hagen2008harmonic}, scientists impact ranking \cite{radicchi2009diffusion,zhou2012quantifying,dorogovtsev2015ranking}, analysis of first-mover advantage for the first publication in career \cite{newman2009first} and highly cited paper prediction \cite{newman2014prediction,zhang2016adawirl}.
%and relationship investigation for different influential node metrics \cite{lu2016h}.
%and topic influence\cite{liu2010mining,tang2009social}.

%Besides Bayesian pairwise ranking, there are many ranking methods in Information Retrieval from a pairwise \cite{burges2005learning,freund2003efficient,joachims2002optimizing} or list-wise perspective \cite{xu2007adarank,wu2010adapting,cao2007learning}.

%The rest of paper is organized as follows. First we report the dataset details and formally define the ARSes predicting problem. Then we reveal a series of factors that contribute to the author's citation increment. With well selected factors, we further design several learning methods for solving the given problem and test their performances via extensive experiments. Finally we give the conclusion of this work.
%\noindent{\textbf{Contribution.}} 

\section*{Results}
\begin{figure}[!t]%[htbp]
\begin{center}
\includegraphics[scale=0.65]{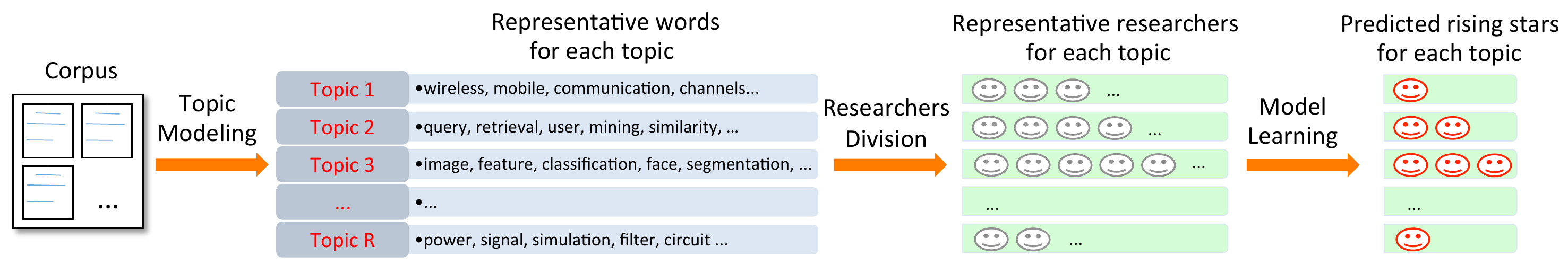}
\caption{Schematic diagram of this work. We divide all young researchers into $R$ groups via topic modeling on their previous publications. Then we extract a series features of each author and design a ranking learning algorithm to predict academic rising stars for each topic/domain.} %The predicted result may provide a useful guidance for faculty hiring or soliciting contribution.}
\label{fig: problem}
\end{center}
\end{figure}

\begin{figure}[!t]%[htbp]
\begin{center}
\includegraphics[scale=0.65]{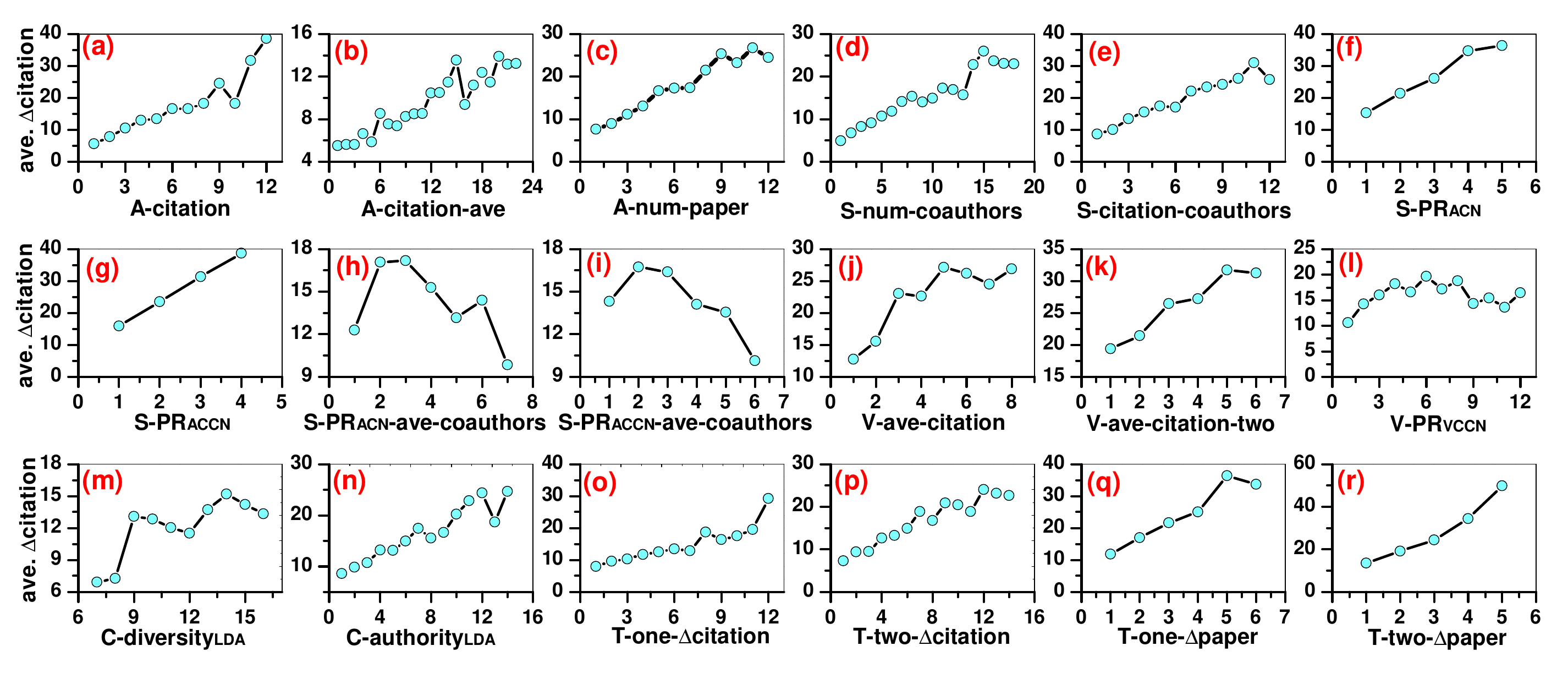}
\caption{The (feature - citation increment) correlations: $x$-$axis$ represents a particular feature's value (e.g., $A$-$citation$, $C$-$diversity_{LDA}$, etc.), and $y$-$axis$ denotes the average citation increment of the authors with the same given feature value. All of the feature values are obtained before current year $t$ (here we set $t=2008$). Each PR score is rescaled by multiplying $10^{6}$.}
\label{fig: feature_citation_correlation}
\end{center}
\end{figure}

\begin{itemize}[leftmargin=0.1in]
\item \textbf{Problem definition.} For ease of representation, Table \ref{tab:notations} lists the notations used throughout the paper. 
%Letter $A$ and $L$ represent the author set and the paper set respectively. Letter $c$ denotes author's citation count and $\Delta  c_{a}$ is researcher $a$'s citation increment in the given time period. The true and predicted impact increment scores for $a$ are symbolized by $s_{a}$ and $\hat{s}_{a}$ respectively.
Considering the various influences, audiences, etc., of different research topics, 
%Symbols used throughout the paper is illustrated in Table 5 (see the \textit{Symbols description} in \textbf{Materials and Methods}). 
we categorize all young researchers $A^{*}$ into $R$ different groups via Latent Dirichlet Allocation (LDA) \cite{blei2003latent} (see \textbf{Researchers division method} in \textbf{Materials and Methods}) and take it as independent prediction task for identifying ARSes for each topic $r$. 
Therefore, we define the problem as:
\textit{Top k\% Academic Rising Stars Prediction - Given the publication corpus $L_{t}$ before the current year $t$ and a set of young researchers $A^{*}$ who publish the first first-author paper at the recent year $t_{1st}$, the task is to predict the fast-rising scholars $A_{r,k}^{*}$ who rank in top $k\%$ in $A_{r}^{*}$ for each topic $r$ according to the citation increments (or impact increment scores) after $\Delta t$ years.}
The schematic diagram of this work is illustrated in Fig. \ref{fig: problem}.
 In this work, the true impact increment score of author $a$ is quantified by citation increment value, i.e., $s_{a}=\Delta  c_{a}$. 
% Let $A^{*}$ represents the set of young scholARSes who publish first firs-author paper at recent timestamp $t_{1st}$ and $A_{r}^{*}$ is the subset of $A^{*}$ belongs to topic $r$. In addition, we use $A_{r,k\%}^{*}$ and $\hat{A}_{r,k\%}^{*}$ to denote the set of true rising stARSes who have top $k\%$ citation increments and the set of predicted rising stARSes who have top $k\%$ predicted impact increment scores for topic $r$, respectively.  
The data extracted from ArnetMiner (see \textbf{Data description} in \textbf{Materials and Methods}), consisted of  1,712,433 authors of computer science, is used for conducting later experiments.

\begin{table*}[htdp]
%\begin{table*}
\begin{center}
\begin{small}
\begin{tabular}{c||l|c||l}
\rowcolor[gray]{.8}
  \hline
   {\bf $Notation$} & {\bf $Definition~and~Description$} &{\bf $Notation$} & {\bf $Definition~and~Description$}\\
   \hline
   \hline
   $A$ & set of researchers, $a$ denotes researcher $a$ & $A^{*}$ & set of researchers who publish first first-author paper at $t_{1st}$\\
   $L$ & set of papers, $l$ represents paper $l$ & $A_{r}^{*}$& set of $A^{*}$ belongs to topic $r$ \\
   $c$ & author' citation number, $\Delta c_{a}$ is $a$'s citation increment& $A_{r,k}^{*}$& set of true rising stars rank in top $k\%$ of topic $r$\\
    $L_{a}$ & the set of papers published by $a$& $\hat{A}_{r,k}^{*}$& set of predicted rising stars rank in top $k\%$ of topic $r$\\
    $s$ & the true impact increment value, $s_{a}$ is $a$'s value&  $\hat{s}$ & the predicted impact increment score, $\hat{s}_{a}$ is $a$'s predicted score\\
     \hline
\end{tabular}
\end{small}
\end{center}
\caption{Notations used throughout the paper.}
\label{tab:notations}
\end{table*}%

\item \textbf{Feature selection.} In order to solve the given problem, 
%we need to detect influences of various features on author's scientific impact increment. 
%Primary analysis on the dataset indicates that 
we formalize the predicted impact increment score $\hat{s}_{a}$ of author $a$ as the combined influence of various factors. %(see \textit{IIRL method} in \textbf{Sec. Materials and Methods}). 
It includes \textit{author}, \textit{social}, \textit{venue}, \textit{content} and \textit{temporal} features of each author, as listed in Table \ref{tab:feature}. In addition, Fig. \ref{fig: feature_citation_correlation} shows the correlation between the average values of citation increments  of $A^{*}$  from year 2008 to year 2012 and the values of a representative feature. There are a group of authors with the same value of the given feature and we compute the average value of citation increments of those authors. To avoid the noise, we only consider the group whose size is larger than 100. We give the description and discussion of each selected feature in the follows.~~~~\noindent{\textbf{Author features.}} The author's impact increment in the future is naturally correlated with their current attributes because: (a) impact of each paper is correlated with author's attributes \cite{castillo2007estimating,yan2012better,petersen2014reputation}. Author's attributes (e.g., previous citation number) influence her/his papers' citation value which in turn further increase their citation numbers. (b) The previous productivity of an author has positive influence on future citation value since she/he has more chances (including self citations) to get citations \cite{bethard2010should}. We extract three author features related with paper number and citation count, namely (1) the number of the author's previous papers, (2) the author's current citation number and (3) the author's current average citation value of previous papers. As what we expect, the results in Fig. \ref{fig: feature_citation_correlation}a to Fig. \ref{fig: feature_citation_correlation}c indicate the positive correlation between the feature's value and author's citation increment: the more papers or the larger citation value an author has, the faster citation increment she/he can expect.
\begin{table*}[htdp]
\begin{center}
\begin{small}
\begin{tabular}{|l||l||l|}
\rowcolor[gray]{.8}
  \hline
   {\bf $Group$} &  {\bf $Feature$} & {\bf $Definition~and~Description$}\\
   \hline
   \hline
   \multirow{3}{*}{$Author$}         & $A$-$num$-$paper$ ($F_{1}$)&  The number of the author's previous publications.    \\
   & $A$-$citation$ ($F_{2}$)& The author's current citation number.\\
        & $A$-${citation}$-$ave$ ($F_{3}$)& The author's current average citation number of previous papers.  \\
        \hline
  \multirow{6}{*}{$Social$} & $S$-$num$-$coauthors$ ($F_{4}$)& The author's previous co-authors number.\\
        & $S$-$citation$-$ave$-$coauthors$ ($F_{5}$)&   The average value of the author's co-authors' citation number.\\
        &$S$-$PR_{ACN}$ ($F_{6}$)&  The author's PR score on ACN.     \\
        &$S$-$PR_{ACCN}$ ($F_{7}$)& The author's PR score on ACCN.     \\
        &$S$-$PR_{ACN}$-$ave$-$coauthors$  ($F_{8}$)& The average value of co-authors' PR scores on ACN.    \\
         &$S$-$PR_{ACCN}$-$ave$-$coauthors$ ($F_{9}$)&  The average value of co-authors' PR scores on ACCN.    \\
         \hline
   \multirow{2}{*}{$Venue$} & $V$-$ave$-$citation$ ($F_{10}$)& The average value of venues' citations of the author's previous papers.\\
         & $V$-$ave$-$citation$-$two$ ($F_{11}$)& The average value of venues' citations of the author's previous papers in the last two years.   \\
          & $V$-$PR_{VCCN}$ ($F_{12}$)& The average PR score of venues of the author's previous papers on VCCN.    \\
                   \hline
   \multirow{2}{*}{$Content$} &$C$-$diversity_{LDA}$ ($F_{13}$)&  The author's diversity value.   \\
         & $C$-$authority_{LDA}$ ($F_{14}$)&  The author's authority score.   \\
            \hline
    \multirow{3}{*}{$Temporal$} &$T$-$one$-$\Delta citation$ ($F_{15}$)&  The citation increment of the author in one year.   \\
          &$T$-$two$-$\Delta citation$ ($F_{16}$)&  The average citation increment of the author in two years.   \\
           &$T$-$one$-$\Delta paper$ ($F_{17}$)&  The paper addition of the author in one year.  \\
            &$T$-$two$-$\Delta paper$ ($F_{18}$)&  The average paper addition of the author in two years.   \\
     \hline
\end{tabular}
\end{small}
\end{center}
\caption{Feature definition and description. We extract five groups of factors for each author including author, social, venue, content
and temporal features. All of the feature values are obtained before the current year $t$ (here we set $t=2008$). Each PR score is rescaled by multiplying $10^{6}$.}
\label{tab:feature}
\end{table*}% 
~~~~\noindent{\textbf{Social features.}} Social interactions among different researchers may influence an author's citation increment. For example, previous studies \cite{bethard2010should,martin2013coauthorship} demonstrated that researchers tend to cite their co-authors' works. To explore such effect, we extract the weighted collaboration network (ACN) among all authors. In ACN, each edge represents a collaboration relationship
between two authors and the weight of an edge is defined as the corresponding collaboration frequency. Besides,
we assume a widely cited author is an authority researcher who has large impact, and construct authors' citing-cited network (ACCN). Unlike ACN,
the ACCN is a weighted directed network and each link denotes a citing-cited relationship between two authors. 
The PageRank (PR) \cite{page1999pagerank} is used to quantify the authority value. Thus we introduce 6 social attributes of each author including: (1) the number of co-authors, (2) the average value of co-authors' citation counts, (3) the author's PR score on ACN, (4) the author's PR score on ACCN, (5) the average value of co-authors' PR scores on ACN and (6) the average value of co-authors' PR scores on ACCN. 
%Each PR score is rescaled by multiplying $10^{6}$ for ease of computing and representation.
The positive effects of co-authors' number, co-authors' average citation value and the author's PR score on ACN or ACCN have been confirmed by Fig. \ref{fig: feature_citation_correlation}d to Fig. \ref{fig: feature_citation_correlation}g. However, according to Fig. \ref{fig: feature_citation_correlation}h and Fig. \ref{fig: feature_citation_correlation}i, the co-authors' average PR value on ACN and ACCN are shown to have negative relationship with the author's citation increment. It should be correlated with the co-authors group size of each author. For example, Fig. \ref{fig: social_and_venue}a and Fig. \ref{fig: social_and_venue}b visualize all collaborators of $a_{1679259}$ who has relative large citation increment and $a_{1689109}$ who has relative large value of $F_{8}$ as well as relative small citation increment, respectively. Node size characterizes the PR score of each co-author and the smallest size in each figure represents around 0.7 score. We can find that $a_{1679259}$ has a large amount of collaborators (like students) and most of them have small PR score, thus the average value of co-authors' PR scores ($F_{8}$) becomes small. Conversely, $a_{1689109}$ tends to collaborate with few influential authors (like advisors) and the $F_{8}$ score of $a_{1689109}$ is relative large due to the small size of co-authors group.~~~~\noindent{\textbf{Venue features.}} Due to various reputations, audiences, etc., different venues have different impacts. In general, good publications in top venues tend to attract more attentions than others. For instance, in the \emph{Network Science} research, researchers pay much more attention to papers published in \emph{Nature} and \emph{Science}. To quantify the venue's influence, we compute the average citation value of all papers in a venue and name it as a venue's citation. Besides, we construct the weighted directed venues' citing-cited networks (VCCN) of all venues to measure the authority of each venue. Each edge represents a citing-cited link between two venues and the weight denotes the citing frequency. Similar to ACN and ACCN, the PR score is used to quantify the venue's authority value. Three venue features of each author are extracted: (1) the average value of venue's citations of the author's previous publications, (2) the average value of venue's citations of the author's previous publications in the last two years and (3) the average PR score of the venues of the author's previous publications on VCCN. The first and second factors positively affect on author's citation increment, as reported by Fig. \ref{fig: feature_citation_correlation}j and Fig. \ref{fig: feature_citation_correlation}k. As for the last factor, it has no strong correlation with author's citation increment. The different corpus sizes of different venues may attribute to this result. Fig. \ref{fig: social_and_venue}c and Fig. \ref{fig: social_and_venue}d show the correlation between average citation increments/total citation increments of all papers and the papers number for each venue. There are a group of venues with the same corpus size and we compute the average value of these venues' total or average citation increments. We consider the group whose size is larger than 10 in order to avoid noise. According to the Fig. \ref{fig: social_and_venue}c and Fig. \ref{fig: social_and_venue}d, we know that venues with larger corpus sizes tend to have larger total citation increments as well as larger PR scores while there is no strong correlation between publication size and average citation increment of each venue. That is to say, some authors published papers in venues with large corpus sizes but small average citation increments, so that the venues have larger PR score and total citation increments but the authors have relatively small citation increments.~~~~\noindent{\textbf{Content features.}} Another factor may affect the author's citation increment is the paper's content. As a popular method for content analysis,
topic modeling is useful for predicting paper's impact \cite{yan2012better,dong2015will}. In problem definition, we use LDA to categorize all researchers into $R$ different groups and it returns the topic distribution of each paper $l \in L_{A^{*}}$. Table \ref{tab:topic-words} reports the five selected representative words of each topic. In general, papers with various topics attract attentions from various research fields. We define topic diversity of author $a$ as the average Shannon entropy \cite{dong2015will} over her/his papers' topic distribution:
\begin{equation*}
diversity(a)=\frac{\sum_{l \in L_{a}} \sum_{r}-p(r|l)\cdot log p(r|l)}{|L_{a}|},
 \end{equation*}
where $p(r|l)$ is the probability distribution over topics $r$ for each paper $l$. Besides the author's diversity, we further define the author's authority over topics as:
\begin{equation*}
authority(a)=\frac{\sum_{r} \sum_{l \in L_{a}}p(r|l) \cdot c_{l}}{\mid  R\mid}.
\end{equation*}
%where ${L_{a}}_{t}$ is $a$'s publications before $t$ and ${c_{l}}_{t}$ is $l$'s citation value at $t$. 
According to Fig. \ref{fig: feature_citation_correlation}m, with the increment of the author's diversity value, the author's citation increment value increases. It indicates that larger diversity attract attention as well as citation from broader areas, which is similar to the correlation between paper's citation and paper's diversity \cite{yan2012better}.
%But when the diversity value goes beyond a certain threshold, the authors's citation increment value decreases with the further increment of diversity value. It indicates that focusing on a few areas may expect large citation increment and too broad interest plays negative role on the researcher' future citation increment. 
Authors with larger authority values tend to have larger citation increment, as confirmed by Fig. \ref{fig: feature_citation_correlation}n. 
\begin{table*}[htdp]
%\begin{table*}
\begin{center}
\begin{small}
\begin{tabular}{c||l|c||l}
\rowcolor[gray]{.8}
  \hline
   {\bf $Topic$} & {\bf $Representative~Words$} &{\bf $Topic$} & {\bf $Representative~Words$}\\
   \hline
   \hline
   1 & optimal, approximation, minimum, matrix, constraints & 6 & services, knowledge, support, application, tool\\
   2 & students, book, science, researchers, engineering & 7& images, features, classification, segmentation, face\\
   3 & management, resources, assessment, strategies, decision& 8& security, scheme, attacks, authentication, solution\\
    4 & wireless, mobile, communication, protocol, channels& 9&parallel, memory, hardware, architecture, distributed\\
    5 & query, retrieval, user, mining, similarity&  10 & power, signal, simulation, filter, circuit\\
     \hline
\end{tabular}
\end{small}
\end{center}
\caption{Representative words of each topic $r$ (here we set $R=10$).}
\label{tab:topic-words}
\end{table*}% 
~~~~\noindent{\textbf{Temporal features.}} The ARSes have good trends of citation increment and typically attract attentions from colleagues easily.
Temporal features of publications have been applied to model paper's scientific impact \cite{yan2011citation,dong2015will}. Similarly, the previous increment of the author's paper number or citation can be good indication for the author's future citation increment. Thus we extract 4 temporal features of each author including: (1) the author's citation addition in previous one year, (2) the author's citation addition in previous two years, (3) the author's paper number increment in previous one year and (4) the author's paper number increment in previous two years. All of four temporal features correlate positively with the author's future citation increments, as reported by Fig. \ref{fig: feature_citation_correlation}o to Fig. \ref{fig: feature_citation_correlation}r.

\item \textbf{Experiment results.}
%In this section, we demonstrate the experiment details including the methods and evaluation metric used in this work, the performances of all methods, as well as the analysis and discussion.\\
%\subsection*{Experimental Setup}
%\noindent{\textbf{Methods.}}
%\noindent{\textbf{Evaluation Metric.}}
Inspired by Bayesian preference ranking \cite{rendle2009bpr, rendle2014improving} in recommender systems, we design an impact increment ranking learning (IIRL) algorithm (see \textbf{IIRL method} in \textbf{Materials and Methods}) for ARSes prediction. For the comparison with IIRL, we use three categories of benchmark methods (see \textbf{Benchmark methods} in \textbf{Materials and Methods}) including a series of regression learning algorithms, various ranking learning algorithms of information retrieval and two naive methods: Base-1 and Base-2. For the regression and ranking learning methods, we only report the results of Naive Bayesian (NB) and RankBoost (RankB) since they reach better performance in respective category.  
As a general machine learning task, we divide $A_{r}^{*}$ for topic $r$ into two parts, one for training set ${A_{r,}^{*}}_{train}$ and the other for test set ${A_{r,}^{*}}_{test}$. The model is trained on ${A_{r,}^{*}}_{train}$ and
 we evaluate its performance on ${A_{r}^{*}}_{test}$.
We take the recall accuracy of the top $k\%$ fastest-rising authors in ${A_{r,}^{*}}_{test}$ as the evaluation metric:
\begin{equation*}
Pre@{k\%}=\frac{{|\hat{A}_{r,k}^{*}}_{test}\bigcap {A_{r,k}^{*}}_{test}|}{|{A_{r,k}^{*}}_{test}|}
\end{equation*}
%\noindent{\textbf{Experimental Results.}}
We construct ${A_{r,}^{*}}_{train}$ with 50\% instances of $A_{r}^{*}$ and use it for model training.
The performance is evaluated on the remaining 50\% instances in ${A_{r,}^{*}}_{test}$. 
In this work, the parameters in problem definition are set as: $R=10$, $t=2008$, $t_{1st}=2006$ and $\Delta t=4$. And the parameters in IIRL algorithm are fixed as: $\alpha=0.01$, $\lambda_{\bm{\omega}}=0.01$. We adjust feature value $f$ of each author via log-transform: $f=ln(f+1)$. 
Fig. \ref{fig: performance} reports the performances of different methods for different topics when $k$ = (10, 20). Overall, NB, RankB and IIRL perform much better than Base-1 and Base-2. It indicates that the selected features are effective in predicting ARSes. Our proposed IIRL reaches the best performance, i.e., (0.51, 0.58) average accuracy for all topics, with over 8\% average improvement than all baselines. The IIRL performs best for most topics because: (a) It transforms the regression task to a pairwise classification problem and can achieve good prediction accuracy for non-linear (power-law like) author's citation increment distribution. (b) The posterior probability of Bayesian ranking model well captures the uncertainty of the future ranking orders of authors' citation increment with the knowledge of current ranking.
We choose IIRL as the primary predicator to examine the following analytical experiments.
\end{itemize}

\section*{Analyses}

\begin{figure}[!t]%[htbp]
\begin{center}
\includegraphics[scale=0.3]{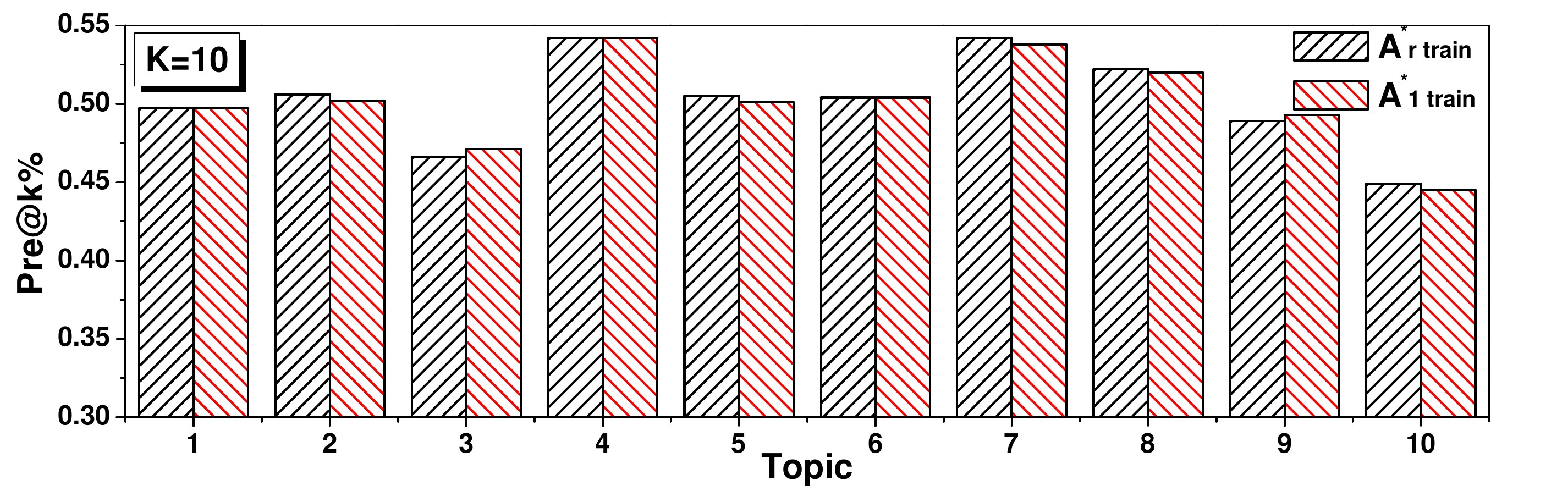}
\includegraphics[scale=0.3]{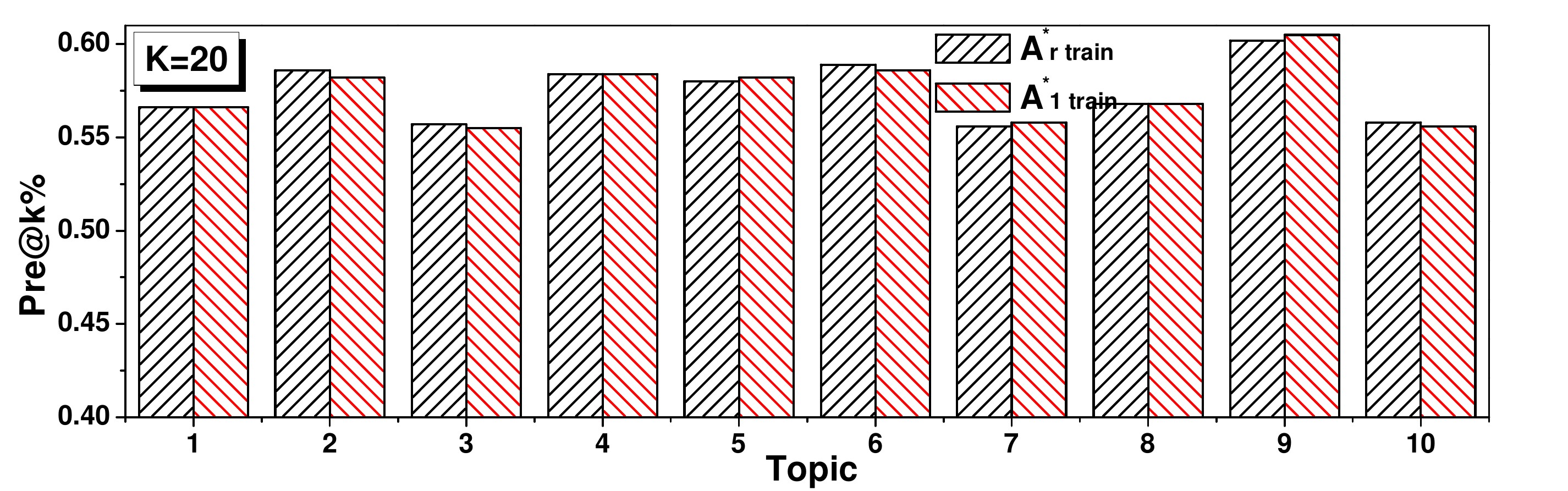}
\caption{Prediction results with two training instances settings for each topic. $Pre@k\%$ of model learned by ${A_{r}^{*}}_{,train}$ and ${A_{\hat{r}}^{*}}_{,train}$ (here we set $\hat{r}=1$) are close to each other for all topics.}
\label{fig: all_topics}
\end{center}
\end{figure}

\begin{itemize}[leftmargin=0.1in]
\item \textbf{Model of all topics.} In problem definition, we divide $A^{*}$ into different groups $A^{*}_{r}$ in respect to various research topics, and take the ARSes prediction as an independent task for each topic $r$. An intuitive question is whether the rising stars prediction models follow the similar pattern for different topics? In other words, can we use the model learned by training instances of one given topic $\hat{r}$ to predict the rising stars belong to other topics $r\neq \hat{r}$? To reveal such problem, we conduct experiment to compare the prediction accuracies of two training settings for each topic: one is that we predict the ${A_{r,k}^{*}}_{test}$ for each topic $r$ by using ${A_{r,}^{*}}_{train}$ independently, and the other is that we identify ${A_{r,k}^{*}}_{test}$ by utilizing ${A_{\hat{r,}}^{*}}_{train}$ of one given topic $\hat{r}$. We randomly choose $\hat{r}=1$ and report the $Pre@k\%$ of IIRL for such two cases in Fig. \ref{fig: all_topics}. According to the result, we can find that $Pre@k\%$ of model learned by ${A_{r,}^{*}}_{train}$ and ${A_{\hat{r,}}^{*}}_{train}$ are close to each other for each topic. It indicates that the rising stars prediction models of different topics follow the similar pattern and we can use the model learned by training instances of one topic to predict the rising stars of any other topics.

\begin{table*}[htdp]
\begin{center}
%\begin{small}
\scriptsize
\begin{tabular}{|c||c|c|c|c|c||c|c|c|c|c||c|}
%\rowcolor[gray]{.8}
  \hline
  \rowcolor[gray]{.8}
  \hline
   \cellcolor[gray]{.8}{\bf $+/-$} & \multicolumn{5}{c||}{\bf $+$} & \multicolumn{5}{c||}{\bf $-$} & \multicolumn{1}{c|}{\bf $$} \\
  \hline
    \rowcolor[gray]{.8}
   \cellcolor[gray]{.8}{\bf $Feature$} & {\bf $Author$}&{\bf $Social$} & {\bf $Venue$}&{\bf $Content$}& {\bf $Temporal$}&{\bf $Author$} & {\bf $Social$}&{\bf $Venue$}&{\bf $Content$}&{\bf $Temporal$}&{\bf $All$}\\
  \hline
  \cellcolor[gray]{.8}{Topic 1}&0.417 & 0.394& 0.246& 0.314& \textbf{0.463}& 0.469& 0.440& 0.480& 0.463& \textbf{0.429}& 0.497\\
   \hline
  \cellcolor[gray]{.8}{Topic 2}&0.420 & \textbf{0.500}& 0.346& 0.370& 0.420& 0.469& \textbf{0.444}& 0.469& 0.469& 0.490& 0.506\\
   \hline
     \cellcolor[gray]{.8}{Topic 3}&0.433& 0.429& 0.290& 0.349& \textbf{0.441}&0.454& 0.466& \textbf{0.450}& 0.454& 0.454& 0.466\\
   \hline
     \cellcolor[gray]{.8}{Topic 4}&0.421 & 0.421& 0.252& 0.336& \textbf{0.487}& 0.529& 0.486& 0.486& 0.481& \textbf{0.479}& 0.542\\
   \hline
        \cellcolor[gray]{.8}{Topic 5}&0.464 & 0.448& 0.328& 0.344& \textbf{0.486}& 0.497& 0.492& 0.486& 0.486& \textbf{0.481}& 0.505\\
   \hline
        \cellcolor[gray]{.8}{Topic 6}&0.449 & \textbf{0.463}& 0.323& 0.354& 0.449& \textbf{0.468}& 0.471& 0.483& 0.490& 0.498& 0.504\\
   \hline
        \cellcolor[gray]{.8}{Topic 7}&0.470& 0.462& 0.311& 0.318& \textbf{0.487}&0.515&  \textbf{0.500}& 0.523& 0.530& 0.515& 0.542\\
   \hline
        \cellcolor[gray]{.8}{Topic 8}&\textbf{0.482} & 0.438& 0.313& 0.348& \textbf{0.482}& 0.509& 0.491& 0.509& 0.509& \textbf{0.482}& 0.522\\
   \hline
        \cellcolor[gray]{.8}{Topic 9}&0.395 & \textbf{0.447}& 0.217& 0.289& 0.434&0.461&  \textbf{0.434}& 0.454& 0.461& 0.461& 0.489\\
   \hline
        \cellcolor[gray]{.8}{Topic 10}&0.342 & 0.383& 0.336& 0.308& \textbf{0.425}& 0.445& 0.425& 0.445& 0.438& \textbf{0.397}& 0.449\\
   \hline
\end{tabular}
%\end{small}
\end{center}
\caption{Contribution analysis of different groups features on different topics.
Temporal features are most influential. Author and social features also have strong effect. Venue features are shown to be least significant.}
\label{tab:performance comparison}
\end{table*}%

\item \textbf{Feature contribution.} We examine the contributions of five groups features by two ways: (1) \textbf{\emph{+Feature.}} Keep only one group of features for model training. (2) \textbf{\emph{-Feature.}} Remove the selected group of features and use the remaining groups of features to train model. The $Pre@k\%~(k=10)$ of IIRL with different feature settings are reported in Table \ref{tab:performance comparison}. According to the result, the temporal features are best indictions for the rising stars prediction. The IIRL keeps over 90\% accuracy with only temporal features and removing them leads to below 0.47 average accuracy. The temporal features capture the rising-trend of each researcher in previous few years. They have strong correlation with future rising-trend and lead to ever-growing impact. Author and social features also have strong influence on prediction result. Meanwhile, the venue features are confirmed to be least significant. Removing them results small loss and using only those features has the poor performance with below 0.29 average accuracy. It indicates that the papers' impacts of rising stars are highly depend on the author's attributes and paper's content quality, regardless of venue level.

\item \textbf{Case study of feature influence.} In order to illustrate the influences of different factors in details, we conduct a case study on a random selected topic $r$ ($r=3$). Table \ref{tab:case study1} reports the top $k\%$ ($k=1$) true rising stars and the predictions by IIRL and Base-2. IIRL correctly predicts 8 rising-stars (with blue color) while Base-2 only has 6 corrected results (with red color). We report the feature information of corrected predictions by IIRL in Table \ref{tab:case study2}, of which the authors with red label are wrong predictions by Base-2. According to the table, $a_{202883}$, $a_{446108}$ and $a_{353968}$ have relative small value of $F_{16}$ ($T$-$two$-$\Delta citation$, with red label) thus Base-2 can not predict them correctly. However, besides temporal features, author and social features are also proved to be influential for the prediction model. For the comparison, we show the medium value of those features of the true rising stars, i.e., the feature value of $a_{mid}$, in the last row. Accordingly, $a_{202883}$, $a_{446108}$ and $a_{353968}$ have relative large values (larger than $a_{mid}$'s value, with blue label) for most of author or social features which affect positively on author's citation increment, and can be correctly predicted by IIRL. Therefore, IIRL leverages various effective factors well and reaches the best performance in ARSes prediction.

\begin{table*}[htdp]
\begin{center}
%\begin{small}
\scriptsize
\begin{tabular}{|>{\columncolor[gray]{0.8}}c||c|}
%\rowcolor[gray]{.8}
  \hline
  \rowcolor[gray]{.8}
  \hline
  % \cellcolor[gray]{.8}{\bf $+/-$} & \multicolumn{5}{c||}{\bf $+$} & \multicolumn{5}{c||}{\bf $-$} & \multicolumn{1}{c|}{\bf $$} \\
 % \hline
{\bf $Method$} & {\bf $Rising~Stars~Id$}\\
  \hline
&423267, 717380, 202883, 1569100, 913009, 1559671, 661842, 902157, 404148, 77052,1353968\\
   \cellcolor[gray]{.8}\multirow{-2}{*}{$True$}& 1438956, 446108, 405996, 870598, 210049, 730385, 802868, 576483, 1342815, 475987, 793285 \\
  \hline
    %\rowcolor{cyan}
  &842488, 1488277, \textcolor{cyan}{202883}, \textcolor{cyan}{1095336}, \textcolor{cyan}{210049}, 503366, \textcolor{cyan}{404148},\textcolor{cyan}{423267}, 1535014, 852180, \textcolor{cyan}{770521}\\
   \cellcolor[gray]{.8}\multirow{-2}{*}{$IIRL$}&405641, 1694381, 1097519, 903059, 417190, 1173298, \textcolor{cyan}{446108},1488661, \textcolor{cyan}{353968}, 12304, 76302 \\
   \hline
&842488, 388726, \textcolor{red}{210049}, 1488661, \textcolor{red}{423267}, 1165827, 1488277, \textcolor{red}{770521},1660245, 1327670, \textcolor{red}{793285}\\
   \cellcolor[gray]{.8}\multirow{-2}{*}{$Base$-2}&\textcolor{red}{404148}, 647788, 417190, 942512, 969548, \textcolor{red}{1095336}, 324600, 1535014, 1375805, 12304, 503366\\
   \hline
\end{tabular}
%\end{small}
\end{center}
\caption{The top $k\%~(k=1)$ true rising stars of topic 3 and the corrected predictions by IIRL and Base-2. IIRL correctly predicts 8 rising-stars while Base-2 only have 6 corrected results.}
\label{tab:case study1}
\end{table*}%

\begin{table*}[htdp]
\begin{center}
%\begin{small}
\scriptsize
\begin{tabular}{|c||c|c|c||c|c|c|c|c|c||c|c|c||c|c||c|c|c|c|}
%\rowcolor[gray]{.8}
  \hline
  \rowcolor[gray]{.8}
  \hline
  % \cellcolor[gray]{.8}{\bf $+/-$} & \multicolumn{5}{c||}{\bf $+$} & \multicolumn{5}{c||}{\bf $-$} & \multicolumn{1}{c|}{\bf $$} \\
 % \hline
   \cellcolor[gray]{.8}{\bf $Feature$} & {\bf $F_{1}$}&{\bf $F_{2}$} & {\bf $F_{3}$}&{\bf $F_{4}$}& {\bf $F_{5}$}&{\bf $F_{6}$} & {\bf $F_{7}$}&{\bf $F_{8}$}&{\bf $F_{9}$}&{\bf $F_{10}$}&{\bf $F_{11}$}&{\bf $F_{12}$}&{\bf $F_{13}$}&{\bf $F_{14}$}&{\bf $F_{15}$}&{\bf $F_{16}$}&{\bf $F_{17}$}&{\bf $F_{18}$}\\
  \hline
  \cellcolor{red}{$a_{202883}$}&  \cellcolor{cyan}{12}& \cellcolor{cyan}{134}& 11.2 &5 & \cellcolor{cyan}{51.4} &  \cellcolor{cyan}{8.1}&  \cellcolor{cyan}{9.1}& 2.5&2.0 &2.6&1.2 &  3.3& 1.6& 41&  35&   \cellcolor{red}{39.0}& 7& 5.0\\
  \hline
    \cellcolor[gray]{.8}{$a_{1095336}$}& 30& 365& 12.2 &16 &26.1 &4.1 &3.1 &1.5 & 1.0&3.4&1.6 &  5.1& 1.5& 99& 59& 68.0& 8& 5.5\\
   \hline
     \cellcolor[gray]{.8}{$a_{210049}$}& 6& 319& 53.2& 13 & 24.5& 1.7& 1.7& 4.0& 3.7&3.4&2.7 &  5.2& 1.4& 58& 94& 99.0& 2&2.0\\
   \hline
    %\rowcolor{cyan}
  \cellcolor[gray]{.8}{$a_{404148}$}&20& 270& 13.5 & 14 & 19.3& 3.8& 3.2&3.0 &2.6&3.3&3.8 & 9.3& 1.3& 49& 36&76.0&3& 5.0\\
   \hline
        \cellcolor[gray]{.8}{$a_{423267}$}& 21& 209& 10.0 & 13& 16.1& 2.4& 2.7&2.3 & 1.9&3.9&4.2 & 6.7& 1.7& 63& 79& 94.5& 9& 8.0\\
   \hline
     \cellcolor[gray]{.8}{$a_{770521}$}& 11& 176& 16.0 & 18& 30.4& 1.4& 1.3&4.0 & 2.9&9.1&8.1 &  5.5& 1.5& 12& 109& 78.5& 5& 3.5\\
   \hline
       \cellcolor{red}{$a_{446108}$}&  \cellcolor{cyan}{18}&  \cellcolor{cyan}{144}& 8.0 & \cellcolor{cyan}{32} &6.7 & \cellcolor{cyan}{4.8} &  \cellcolor{cyan}{4.6}&2.1 & 1.8&2.1&0.6 &  1.7& 1.5&  41&  55&   \cellcolor{red}{38.0}& 5& 5.0\\
   \hline
   \cellcolor{red}{$a_{353968}$}& \cellcolor{cyan}{9}&  \cellcolor{cyan}{188}& \cellcolor{cyan}{20.9}& 10& \cellcolor{cyan}{23.2} &  \cellcolor{cyan}{2.8}& \cellcolor{cyan}{2.5}& 2.3& 1.8& 2.5&3.4 &  1.6& 1.7& 27&  48&   \cellcolor{red}{48.5}& 2& 3.5\\
   \hline
          \cellcolor[gray]{.8}{$a_{mid}$}& 9& 134& 13.3 &14&16.1&2.0 &1.7 &-- &-- &-- & --&--&--&-- & --& --& --& --\\
   \hline
\end{tabular}
%\end{small}
\end{center}
\caption{Feature information of corrected predictions by IIRL. $a_{202883}$, $a_{446108}$ and $a_{353968}$ have relative small value of $T$-$two$-$\Delta citation$ and relative large values for most of author or social features.}
\label{tab:case study2}
\end{table*}%
\section*{Conclusion  \& Discussion}
In this work, we formalize a new problem for predicting the top $k\%$ fastest rising young researchers in citation number. To solve the problem, we explore a series of factors that can drive a young researcher to be ARSes and design a novel impact increment ranking learning algorithm to effectively predict the ARSes of each research topic/domain. Our method helps identify ARSes in advance, which may offer useful guidance for research community like young faculty hiring of university. Our further analysis demonstrates that the prediction models for different research topics follow the similar pattern. We also conclude that the temporal features are the best indication for rising stars prediction, while the venue features are little relevant. Despite the current satisfied outcome, there are still some promising future works. For example, there is an improvement space for the prediction accuracy of IIRL method although it outperforms all benchmark methods. An optimized IIRL algorithm which trains the prediction model by integrating the current ranking information of authors in each iteration step, may be a good choice for improving performance. 
\end{itemize}
%\noindent{\textbf{Information spreading mechanism.}}

%\section*{Conclusion}

\section*{Materials \& Methods}
\begin{itemize}[leftmargin=0.1in]
\item \textbf{Data description.}
We use a large real-world dataset from ArnetMiner which is a well known online service for academic search and analysis. The dataset contains 1,712,433 authors and 2,092,356 papers from major computer science venues for more than 50 years (from 1960 to 2014). Each paper contains content information on the title, authorship, abstract, publication time, publication venue and references. In total, we extract 4,258,615 collaboration relationships among authors and 8,024,869 citation relationships among papers from the dataset.

\item \textbf{Researchers division method.}
Considering the various influences, audiences, etc., of different research topics/domains, we should take it as independent prediction task for each topic. As a widely used method for topic modeling, we use Latent Dirichlet Allocation\footnote{\url{http://radimrehurek.com/gensim/}} (LDA) to categorize the corpus into $R$ different research topics. We run a $R$-topics LDA on the title and abstract of $L_{A^{*}}$ and it returns the probability distribution $p(r|l)$ over topic $r$ for each paper $l \in L_{a}$ of each researcher $a \in A^{*}$. The topic probability distribution $p(r|a)$ for $a$ is defined as the summation over $p(r|l)$ of each $l\in L_{a}$: $p(r|a)=\sum_{l \in L_{a}} p(r|l)$.
Note that most of researchers' works cover several different topics, we divide all researchers of $A^{*}$ into $R$ groups and each $a\in A^{*}$ belongs to $m~(m=3)$ different groups according to the top $m$ values of $p(r|a)$ for all topics. 

\item \textbf{Benchmark methods.}
For the comparison with IIRL, we use three categories of baseline methods including: \emph{Category $\uppercase\expandafter{\romannumeral1}$}. A series of regression learning methods\footnote{\url{http://scikit-learn.org/}} which predict the rising stars according to the predicted impact increment score. It contains Logistic Regression, Naive Bayesian (NB), Random Forest, Support Vector Machine and Gradient Descent Boosting Tree. We only report the results of NB because it has better performance.  \emph{Category $\uppercase\expandafter{\romannumeral2}$}. Various ranking algorithms\footnote{\url{http://people.cs.umass.edu/~vdang/ranklib.html}} in information retrieval, which predict the rising stars according to the predicted rank orders of impact increment score. It includes RankNet, RankBoost (RankB), AdaRank and Coordinate Ascent. With proper parameters, RankB achieves better performance and we report it. \emph{Category $\uppercase\expandafter{\romannumeral3}$}. Two naive methods including: Base-1 which predicts the rising stars according to current author's citation value and Base-2 which ranks the authors by using their average citation increments in previous two years. 
%Besides available regression and ranking methods, we design a novel impact increment ranking learning (\emph{IIRL}) algorithm, which is inspired by \emph{Bayesian~Preference~Ranking} \cite{rendle2009bpr,rendle2014improving} in recommender systems. The mathematical deduction for \emph{IIRL} is illustrated in \textit{IIRL method} of \textbf{Sec. Materials and Methods}. We compare the prediction results of our methods with \emph{Base-1} method which predicts the rising stARSes according to current author's citation value and \emph{Base-2} method which ranks the authors by using their average citation increment of previous two years. 

\item \textbf{Impact Increment Ranking Learning (IIRL) method.}
In this work, inspired by Bayesian preference ranking \cite{rendle2009bpr, rendle2014improving} in recommender systems, we design an impact increment ranking learning (IIRL) method for ARSes prediction.
%Inspired by previous works[], we introduce the Bayesian pairwise ranking model to solve the problem.
Let $(a_{i},a_{j})\in T_{r}$ denote an author pair of topic $r$.
In order to capture impact increment rankings of different authors categorized in the same topic,
we maximize the posterior probability with parameter $\bm{\omega}$: 
\begin{equation*}
p(\bm{\omega}|>_{r})\propto p(>_{r}|\bm{\omega})\cdot p(\bm{\omega})
\end{equation*}
where notation $>_{r}=\{a_{i}>_{r}a_{j}:((a_{i},a_{j})\in T_{r}) \cap (s_{a_{i}} > s_{a_{j}})\}$ represents the pairwise structure
of topic $r$ and $p(\bm{\omega})$ is the prior probability.
We take citation increment in the given time period as the true impact increment score, i.e., $s_{a_{i}}=\Delta c_{a_{i}}$.
Let set ${T_{>}}_{r}$ represent all instances in $>_{r}$ and set ${T_{\leq}}_{r}$ consists of the remaining cases not included in ${T_{>}}_{r}$. In general,
we assume that each case in $>_{r}$ is independent. The likelihood function can be written as a product of single density for all cases in $>_{r}$:
\begin{small}
\begin{equation*}
%\begin{split}
p(>_{r}|\bm{\omega})=\prod_{(a_{i},a_{j})\in T_{r}}p(a_{i}>_{r}a_{j}|\bm{\omega})^{\delta((a_{i},a_{j})\in {T_{>}}_{r})}\\
\cdot (1-p(a_{i}>_{r}a_{j}|\bm{\omega}))^{\delta((a_{i},a_{j})\in {T_{\leq}}_{r})}
%\end{split}
\end{equation*}
\end{small}
With antisymmetric nature of $>_{a}$, the log form of the objective becomes:
\begin{small}
\begin{equation*}
ln~p(\bm{\omega}|>_{r})=ln~\prod_{(a_{i},a_{j})\in {T_{>}}_{r}}p(a_{i}>_{r}a_{j}|\bm{\omega})\cdot p(\bm{\omega})\\
=\sum_{(a_{i},a_{j})\in {T_{>}}_{r}}ln~p(a_{i}>_{r}a_{j}|\bm{\omega})+ln~p(\bm{\omega})
\end{equation*}
\end{small}
Let $p(a_{i}>_{r}a_{j}|\bm{\omega})= \sigma(d_{(a_{i},a_{j})\in {T_{>}}_{r}}(\bm{\omega}))$,
where $\sigma$ is the logistic function: $\sigma(x)=\frac{1}{1+e^{-x}}$, and $d_{(a_{i},a_{j})\in {T_{>}}_{r}}(\bm{\omega})$ measures
 the impact increments difference between $a_{i}$ and $a_{j}$. The predicted impact increment score $\hat{s}_{a_{i}}$ of $a_{i}$
 based on academic features $\bm{f}$ (described in \textbf{Feature selection}) extracted from the corpus is formulated as $\hat{s}_{a_{i}}=\sum_{k=1}^{K} \bm{\omega}_{k}\cdot f_{ik}$,
where $f_{ik}$ represents the $k$-$th$ factor of $a_{i}$ and $K$ is the number of all factors. Intuitively,
we define $d_{(a_{i},a_{j})\in {T_{>}}_{r}}(\bm{\omega}) = \hat{s}_{a_{i}}-\hat{s}_{a_{j}}$ and parameter prior distribution as $\bm{\omega} \sim \mathcal{N}(\bm{0},\lambda_{\bm{\omega}}\bm{I})$. Therefore the objective becomes:
\begin{small}
\begin{equation*}
IIRL_{objective} \equiv \sum_{(a_{i},a_{j})\in {T_{>}}_{r}}ln~\sigma (\hat{s}_{a_{i}}-\hat{s}_{a_{j}})-\lambda_{\bm{\omega}}\cdot \|\bm{\omega}\|^{2}
\end{equation*}
\end{small}
where $\lambda_{\bm{\omega}}$ is the model specific regularization parameter.
The IIRL objective function is differentiable thus we use stochastic gradient descent \cite{zhang2004solving} for maximization.
Specifically, IIRL random initialize the parameter $\bm{\omega}$ according to
$\bm{\omega} \sim \mathcal{N}(\bm{0},\lambda_{\bm{\omega}}\bm{I})$, then iteratively traverses each pairwise structure $(a_{i},a_{j})\in {T_{>}}_{r}$ and updates $\bm{\omega}$ by following rule until it meets the stop criterion:
\begin{small}
\begin{equation*}
%\begin{split}
\bm{\omega} \leftarrow \bm{\omega} + \alpha \cdot \frac{\partial~IIRL_{objective}}{\partial \bm{\omega}} \\
= \bm{\omega} + \alpha \cdot \{\frac{e^{-(\hat{s}_{a_{i}}-\hat{s}_{a_{j}})}}{(1+e^{-(\hat{s}_{a_{i}}-\hat{s}_{a_{j}})})}\cdot
\frac{\partial (\hat{s}_{a_{i}}-\hat{s}_{a_{j}})}{\partial \omega}-\lambda_{\bm{\omega}}\cdot \bm{\omega}\}
%\end{split}
\end{equation*}
\end{small}
where $\alpha$ is the learning rate.

\end{itemize}
%\bibliographystyle{naturemag}
%\bibliography{E:/Papers/Auxiliary/Bibliography}

%\begin{thebibliography}{10}
%\expandafter\ifx\csname url\endcsname\relax
%  \def\url#1{\texttt{#1}}\fi
%\expandafter\ifx\csname urlprefix\endcsname\relax\def\urlprefix{URL }\fi
%\providecommand{\bibinfo}[2]{#2}
%\providecommand{\eprint}[2][]{\url{#2}}
%***
%\end{thebibliography}
\bibliographystyle{naturemag}
 \bibliography{refs}
%\end{thebibliography}

% \noindent{\textbf{Random Training Instances Sampling.}}
%Generally, IRL iteratively traverses each $(a_{i},a_{j})\in {T_{>}}_{r}$ for parameters optimization. However, it takes long time to train model each round due to the large size of all pairwise structures in ${A_{r}^{*}}_{train}$ for each topic $r$. For example, suppose the number of authors in ${A_{r}^{*}}_{train}$ is 1000, the size of all pairwise structures $M \sim C^{2}_{1000}$.
\section*{Acknowledgements}
This work was partially supported by the National Natural Science Foundation of China (No. 11305043, No. 61433014), and the Zhejiang Provincial Natural Science Foundation of China (No. LY14A050001), the EU FP7 Grant 611272 (project GROWTHCOM) and Zhejiang Provincial Qianjiang Talents Project (Grant No. QJC1302001). Chuxu Zhang thanks to the Assistantship of Computer Science Department of Rutgers University.
\vspace{-5mm}
\section*{Author contributions}
\vspace{-3mm}
\noindent{C.Z., C.L., L.Y., Z.K.Z and T.Z. designed the research; C.Z. and C.L. analyzed data and performed the experiment; C.Z., C.L., L.Y., Z.K.Z and T.Z. wrote and revised the manuscript.}

\vspace{-5mm}
\section*{Additional information}
\vspace{-3mm}

%\noindent{{\textbf{Supplementary information}} accompanies this paper at http://www.nature.com/scientificreports}

\noindent{{\textbf{Competing financial interests:}} The authors declare no competing financial interests.}

\vspace{2mm}

%\noindent{{\textbf{License:}} This work is licensed. }

%\vspace{2mm}

%\noindent{\textbf{How to cite this article:} ***}

\clearpage
\newpage

\begin{figure}[!t]%[htbp]
\begin{center}
\includegraphics[scale=0.18]{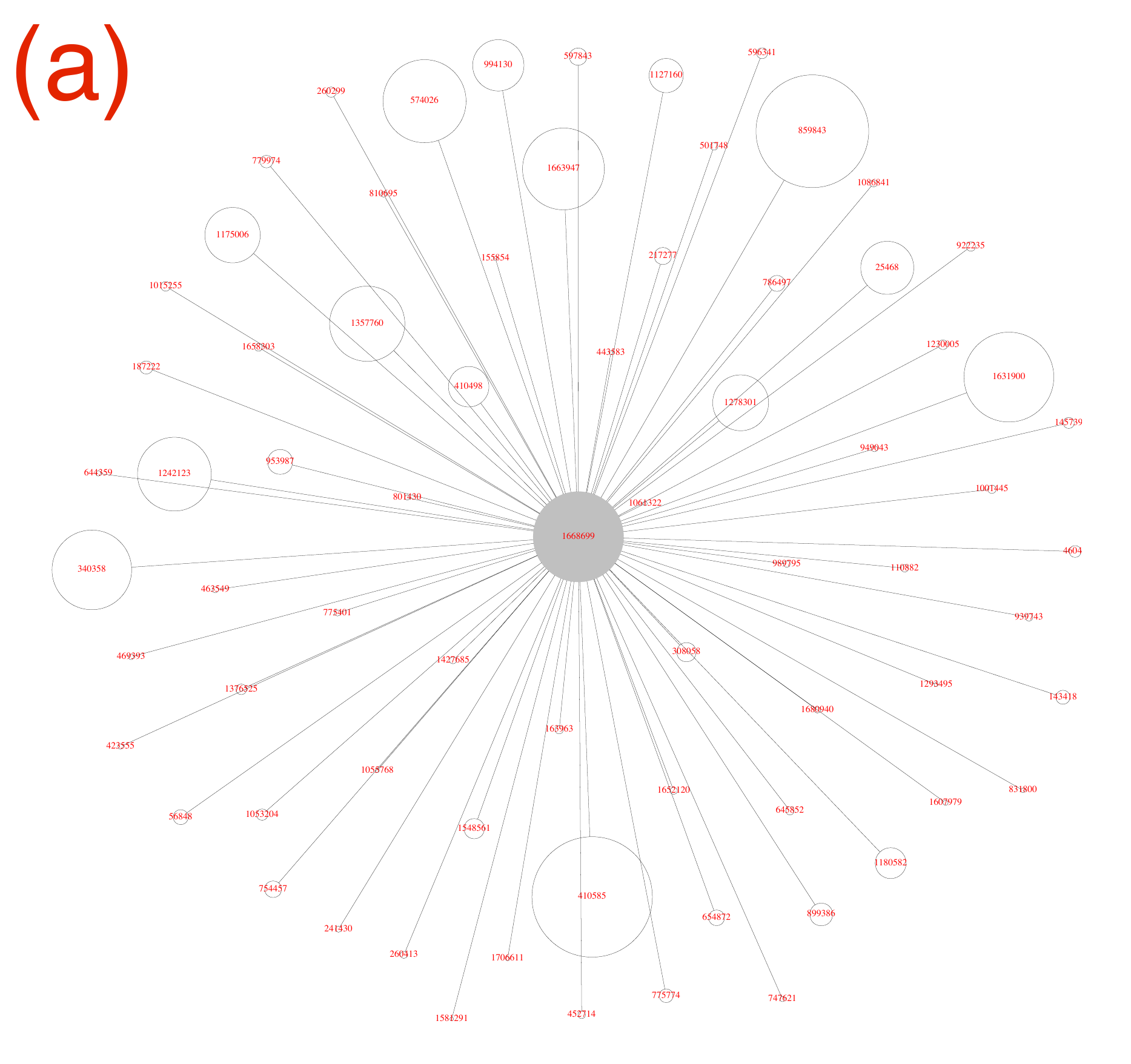}
\includegraphics[scale=0.18]{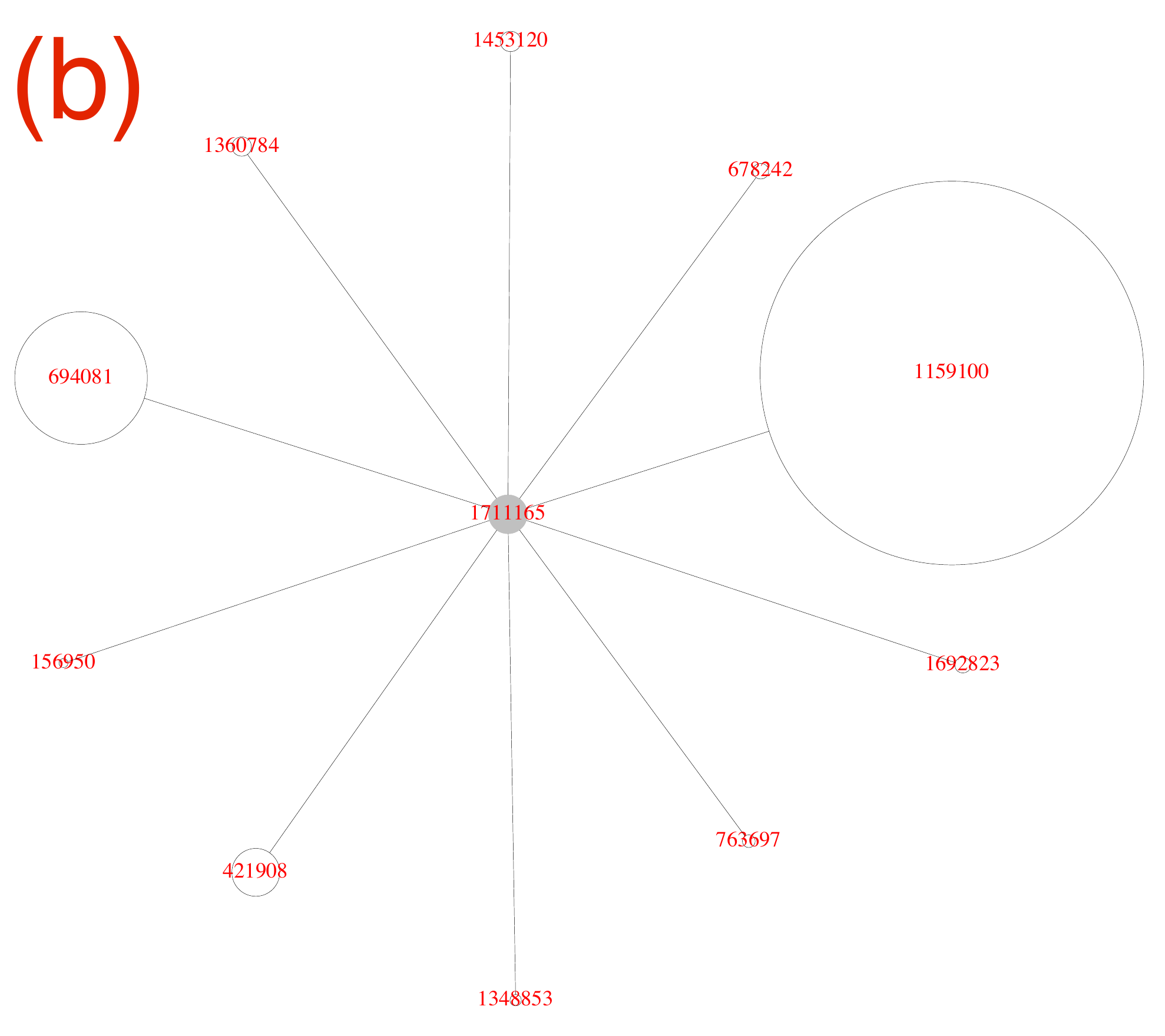}
\includegraphics[scale=0.18]{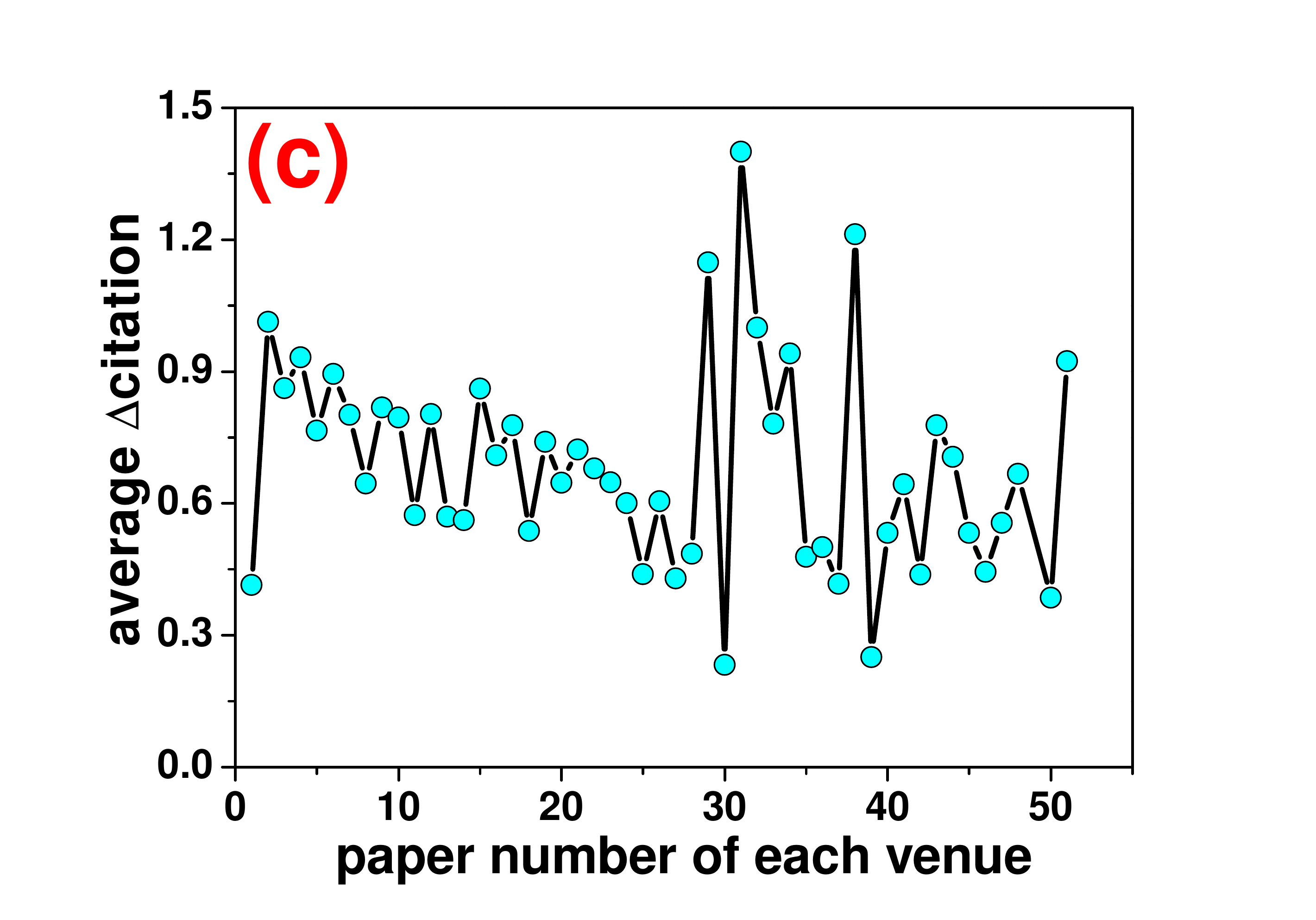}
\includegraphics[scale=0.18]{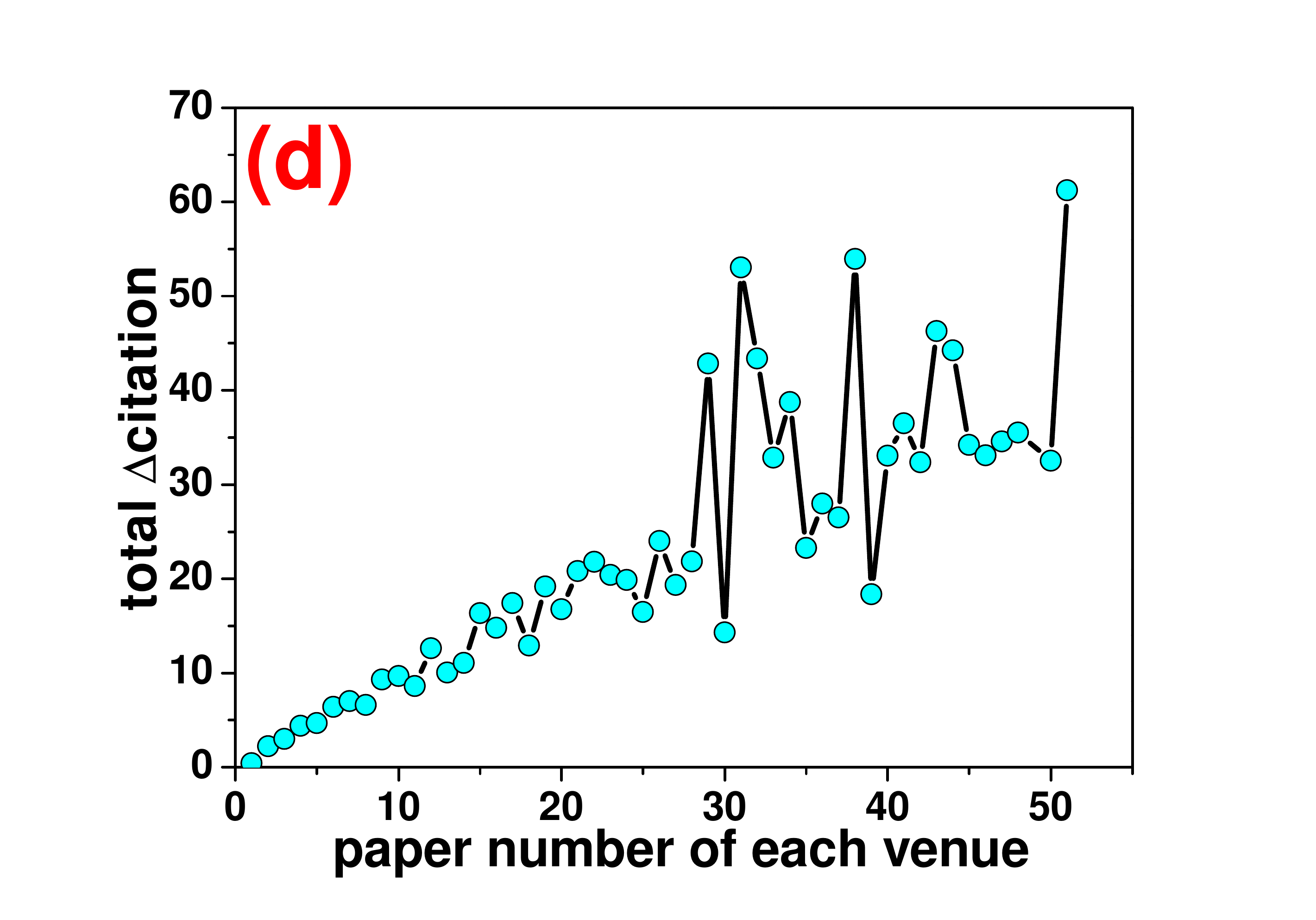}
\caption{((a) Collaborators visualization of $a_{1679259}$. (b) Collaborators visualization of $a_{1689109}$. (c) The correlation between average citation increment and the papers number of each venue. (d) The correlation between total citation increment and the papers number of each venue.}
\label{fig: social_and_venue}
\end{center}
\end{figure}

\begin{figure}[!t]%[htbp]
\begin{center}
\includegraphics[scale=0.6]{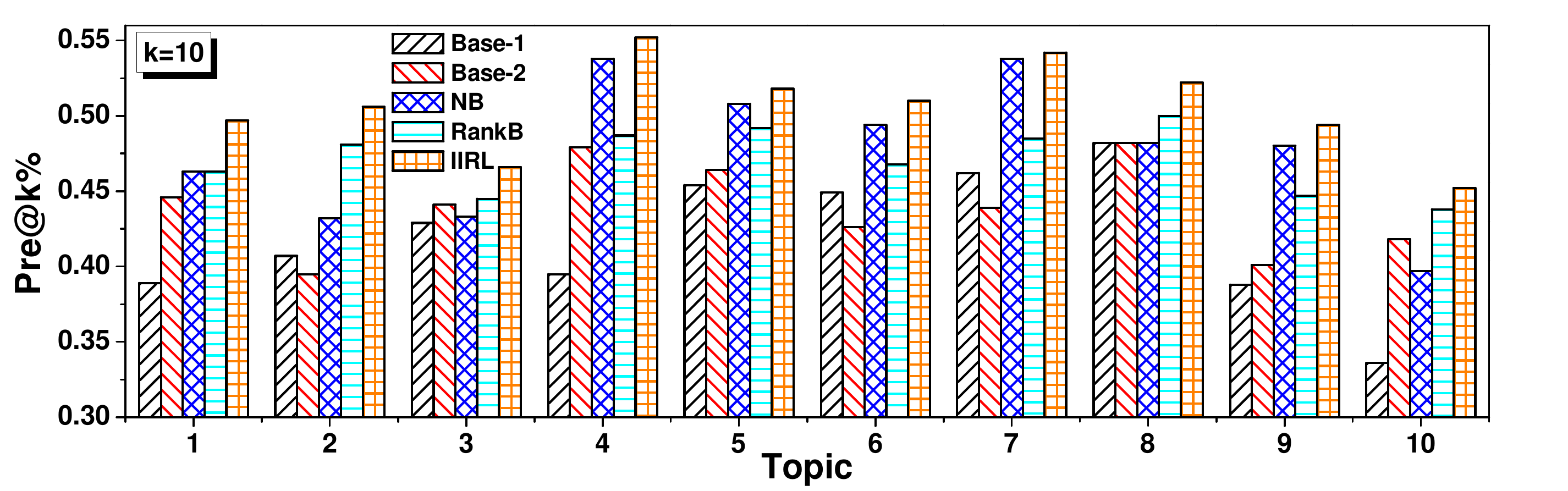}
\includegraphics[scale=0.6]{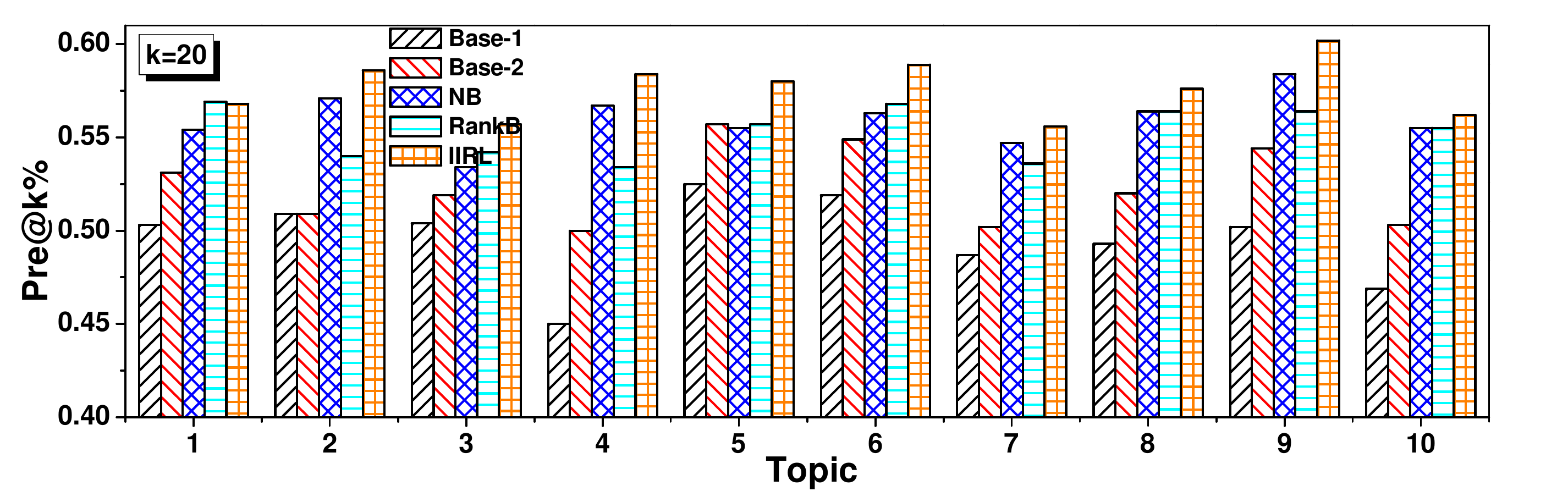}
\caption{Performance comparison of different methods. NB, RankB and IIRL perform much better than Base-1 and Base-2. Our proposed IIRL reaches the best performance for most topics.}
\label{fig: performance}
\end{center}
\end{figure}

\end{document}